\documentclass[namedreferences]{solarphysics}
\usepackage[hyperref,optionalrh,solaromanenum]{spr-sola-addons} 
\usepackage{graphicx}                    
\usepackage{color}                       

\begin{document}

\begin{article}

\begin{opening}

\title{Clarifying Physical Properties of Magnetic Fields in Sunspots}

\author[addressref={1},
email={obridko@mail.ru}]{\inits{V.N.\,}\fnm{V.N.\,}\lnm{Obridko}\orcid{0000-0001-5100-806X}}
\author[addressref={2},
email={mkatsova@mail.ru}]{\inits{M.M.\,}\fnm{M.M.\,}\lnm{Katsova}\orcid{}}
\author[addressref={1,3},corref,
email={sokoloff.dd@gmail.com}]{\inits{D.D.\,}\fnm{D.D.\,}\lnm{Sokoloff}\orcid{0000-0002-3441-0863}}
\author[addressref={1},
email={shelting@izmiran.ru}]{\inits{B.D.\,}\fnm{B.D.\,}\lnm{Shelting}\orcid{}}
\author[addressref={1,2},
email={ilivsh@gmail.com}]{\inits{I.M.\,}\fnm{I.M.\,}\lnm{Livshits}\orcid{0000-0002-8390-013X}}
\runningauthor{V.N. Obridko  et al.}
\runningtitle{Properties of Magnetic Fields in Sunspots}

\address[id={1}]{IZMIRAN,  4 Kaluzhskoe  Shosse, Troitsk, Moscow, 108840,  Russia}
\address[id={2}]{Sternberg State Astronomical Institute, Lomonosov Moscow State University, Universitetskij prosp.13, Moscow, 119991, Russia}
\address[id={3}]{Moscow State University,  Moscow, 119991, Russia}
\begin{abstract}  
We demonstrate that the radial magnetic-field component at the outer boundary of the sunspot penumbra is about 550 Mx\,cm$^{-2}$ independent of the sunspot area and the maximum magnetic field  in the  umbra. The mean magnetic-field intensity in sunspots grows slightly as the sunspot area increases up to 500\,--\,1000 {\it millionth of visual hemisphere} (m.v.h.) and may reach about 900\,--\,2000 Mx\,cm$^{-2}$. The total magnetic flux weakly depends on the maximum field strength in a sunspot and is determined by the spottedness, i.e. the sunspot number and the total sunspot area; however, the relation between the total flux and the sunspot area is substantially nonlinear. We suggest an explicit parametrization for this relation. The contribution of the magnetic flux associated with sunspots to the total magnetic flux is small, not achieving more than 20\,\% even at the maximum of the solar activity.

\end{abstract}

%
\keywords{Sunspots, Magnetic Fields; Sunspots, Penumbra; Sunspots, Umbra}

\end{opening}

%
 \section{Introduction}\label{intr} 

A sunspot is defined as an area of enhanced magnetic field on the solar surface; however, a specific value of the magnetic field strength that determines the sunspot boundary is lacking. Such a definition is important, in particular, because this is the field value at which the heating of the  photosphere  by convection decreases, and a dark feature forms on the solar surface. However, it is very difficult to measure this critical value directly on the jagged boundary or to draw contour lines, especially when we are dealing with high-resolution magnetograms. Moreover, the results of direct measurements for different spots will inevitably depend on individual properties of a given sunspot. Another reason for  determining the mean properties of sunspot magnetic fields is that the series of observation data on sunspot numbers and areas are much longer than the magnetic field time series. The relationship between the mean sunspot magnetic field, on the one hand, and the sunspot area and number, on the other, can allow us to convert the historical sunspot observation data to the magnetic flux.     

The databases of the total sunspot areas (spottedness) cover about 150 years, while the sunspot magnetic-field time series are much shorter. A straightforward idea to convert the spottedness data to the magnetic flux is to use the sunspot magnetic field. However, the solar survey measurements usually give the maximum magnetic field in a given sunspot. So, using these data directly one will inevitably overestimate the sunspot magnetic flux. Besides, there is reason to believe that the mean magnetic field in a sunspot itself may depend on the spottedness.  Finally, the solar magnetic field outside the sunspot also contributes to the total magnetic flux, and the relation of this contribution to that of the sunspots can be nonlinear.     
 
In order to specify the mean value of the magnetic field  in sunspots, we should first introduce the concept of the sunspot magnetic boundary. This concept must be consistent with the traditional notion of the sunspot boundary based on photometric data, which underlays the long-term databases of sunspot areas.
 
For quite a long time, the magnetic field outside sunspots was considered negligible. So, equations were derived, according to which the magnetic field vanishes at the outer boundary of the penumbra  \citep{B42, M53}, and  the dependence of the field intensity on the distance from the center of a symmetric spot was fully determined by the maximum magnetic field at the center. Later, various estimates of $B_0/B_{\rm b} =c$ were adopted (here, $B_0$ is the magnetic field at the sunspot center and $B_{\rm b}$ is the field at the sunspot boundary), in particular, $c=0.5$ \citep{BS69}, $c=0.2$ \citep{W74}, $c=0.163$ \citep{GH81}, and $c=0.607$ \citep{K83}.  

Recently, special attention has been drawn to the threshold value of the magnetic field at the umbra--penumbra boundary  \citep{J11, Jetal15, Jetal17, Jetal18, Setal18, Letal20}. The point is that the umbra--penumbra boundary is where the vertical magnetic field of the umbra is transformed into the  mainly horizontal magnetic field. The vertical magnetic field here is estimated as  1867\,G. This quantity is determined with high accuracy and does not depend on the size and evolution of the sunspot, while the ratio of the vertical and horizontal magnetic fields can vary both along the boundary and during the evolution of a sunspot. According to \cite{Jetal18}, the very existence of this universal quantity seems very important. It can be reasonably interpreted as a quantity that determines the lower field value that is sufficient to saturate the magnetic convection. At this value, the field ceases to be mainly vertical, and penumbral filaments appear with a strong horizontal component. \cite{MM19} explained this result theoretically based on the criterion by \cite{GT66}. They arrived at a conclusion that the umbra--penumbra boundary is where the vertical field is strong enough to increase the effective adiabatic temperature gradient by two orders of magnitude over its non-magnetic value. 

On the other hand, there is some doubt concerning the universal character of the above assessment, as well as  of the precise value of the vertical magnetic field at the umbra--penumbra boundary. In particular, \cite{Eetal21} obtained a substantially smaller estimate (1576\,G) of the quantity under discussion. \cite{Betal18} argued that the boundary was unstable, at least in  decaying sunspots, and they did not find any stable estimate for the vertical magnetic field there.  \cite{Metal16} investigated the development of the Evershed effect in the vicinity of the pores and demonstrated that the situation was very unstable and varied over a time scale of one or three hours. \cite{Loetal20} questioned the very existence of a universal connection between the vertical magnetic field and the umbra--penumbra boundary in sunspots. 

The vertical field at the outer boundary of the penumbra is even more uncertain. Besides the highly contradictory data mentioned above, the most comprehensive study with a Stokes polarimeter was performed for 16 different sunspots by \cite{KM96}. As follows from the figure presented therein, the vertical field component at the outer boundary of the penumbra is about 300\,Mx\,cm$^{-2}$ and the radial component is about 800\,Mx\,cm$^{-2}$. Citing that work, \cite{Setal06}  simply indicate the general range of 500\,--\,1000\,Mx\,cm$^{-2}$.  \cite{Aetal13}, assuming that the maximum field value at the spot center is 3500\,Mx\,cm$^{-2}$ and using a factor of 1/5, obtain the field at the boundary of the umbra equal to 700\,Mx\,cm$^{-2}$. \cite{Hetal21} (with an incorrect reference to \cite{Aetal13}) assume that the field at the outer boundary is 500\,Mx\,cm$^{-2}$ and, multiplying by 5, obtains the sunspot field of 2500\,Mx\,cm$^{-2}$.
\cite{BI11} investigated magnetic fields in two rather regular sunspots to find out that the magnetic-field profile inside a sunspot as a function of the distance to the spot center was rather stable. The vertical component of the magnetic field at the umbra--penumbra boundary represented in the figure in this article is about 1500\,Mx\,cm$^{-2}$ with an uncertainty of about a few hundred Mx\,cm$^{-2}$, whereas at the outer boundary of the penumbra, it is about 500\,Mx\,cm$^{-2}$ with the same uncertainty. 

We see that the above approach remains debatable and various estimates are suggested. However, the existence of a quantitative relation between the boundary of a sunspot including the penumbra and the magnetic-field strength, which determines the appearance of a dark area on the solar surface, looks very reasonable. Since the methods used  to determine the magnetic-field strength and the photometric boundary are basically different, \cite{OS18} suggested finding a threshold, where the distribution of daily areas with the magnetic field larger than this threshold value coincides with the distribution of daily sunspot areas. This threshold field  determines a boundary, which can be referred to as the mean sunspot magnetic boundary.   

The aim of this article is to clarify the following questions

\begin{enumerate}
\item Does the magnetic field at the outer boundary of the penumbra depend on the sunspot area and is there a universal relation between these quantities?

\item Does the magnetic field at the outer boundary of the penumbra depend on the maximum magnetic field in the sunspot umbra and is there a universal relation between these quantities?

\item How does the mean magnetic field change in sunspots of different areas?

\item How does the total magnetic flux from the entire solar disk change and what is the determining factor in this variation --  the total sunspot area or the magnetic field variations in individual spots? 
\end{enumerate}

\section{Methods and Databases}\label{meth}

As stated above, the main objective of our study is to establish a relationship between the sunspot areas measured by photometric methods (including the time when magnetic data did not exist) and the magnetic flux. To do this, we have to determine the magnetic boundary of an area occupied by a photometric spot, which would be consistent with the entire set of observations. This boundary  naturally varies from spot to spot depending on their particular properties. Therefore, we do not study individual spots nor compare their photometric and magnetic maps directly. Such comparisons have been made repeatedly. So, in particular, magnetic-field maps with spot contours drawn on them are shown by \cite{BI11}. The magnetic-field values at the sunspot boundary reported in the mentioned work generally agree with our results presented below.

To accomplish our task, we apply the method of comparing two databases.
We use data on the daily longitudinal magnetic field {\it Helioseismic and Magnetic Imager}  \citep{Setal12, Sch12} onboard the  {\it Solar Dynamics Observatory} (SDO/HMI) \citep{Petal12} for 2375 days from 01 May 2010 to 31 October 2016. The daily sunspot numbers were taken from  {\it World Data Center-Sunspot Index and Long-term Solar Observations} (WDC-SILSO), Royal Observatory of Belgium, Brussels  \textsf{side.oma.be/silso/datafiles} (version 2). The cumulative daily sunspot areas were taken from the NASA Web site \textsf{solarscience.msfc.nasa.gov/greenwch.shtml}.

At present, there are two databases formed of high-resolution observations carried out with similar instruments. These are 
{\it Michelson Doppler Imager}  \citep{Setal95}
onboard the {\it Solar and Heliospheric Observatory} (SOHO/MDI)  continuously measured the Doppler velocity, longitudinal magnetic field, and brightness of the Sun for 15 years up to 12 April 2011 and the enhanced SOHO/HMI  began making its routine observations on 30 April 2010. HMI data include all MDI observables, but with much higher spatial and temporal resolutions, better data quality and a different spectral line. The optical resolution of these instruments is comparable and is 1.17 arcsec and 0.91 arcsec for MDI and HMI, respectively. At the same time, the size of pixels differs essentially. In full-disc observations, it is 1.98 arcsec for MDI and almost four times less (0.505 arcsec) for HMI. A careful pixel-by-pixel comparison of the HMI and MDI signals was performed by \cite{Letal12}. The noise of a single measurement was 10.2\,Mx\,cm$^{-2}$ for the 45-second HMI magnetograms and 26.4 Mx cm$^{-2}$ for the one--minute full--disk MDI magnetograms. The averaging over longer intervals, naturally, somewhat decreases the noise level. The noise in the HMI and MDI line-of-sight magnetic-field synoptic charts appears to be fairly uniform over the entire map. The noise is 2.3\,Mx\,cm$^{-2}$ for HMI charts and 5.0\,Mx\,cm$^{-2}$ for MDI charts. Besides that, the line-of-sight magnetic signal inferred from the calibrated MDI data is greater than that derived from the HMI data by a factor of 1.40.

Here, we have transformed the daily values of the longitudinal magnetic field component into the radial component by dividing by $\cos \theta$, where $\theta$ is the position angle. The area of each pixel was corrected by the same factor to calculate the total magnetic flux in a circle of a given radius for the magnetic fields larger than the given threshold. This yields the relative part of the area occupied by the magnetic field larger than the threshold vale. 
This relative area is expressed in millionths of the visible hemisphere (m.v.h.), as is customary when studying the total sunspot areas.

\begin{figure} 
\centerline{\includegraphics[width=0.5\textwidth,clip=]{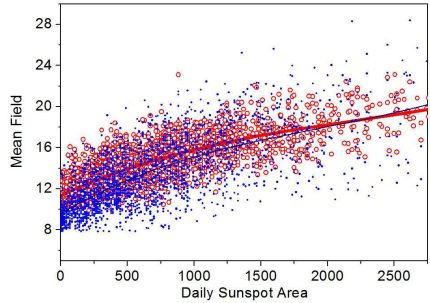}}
\caption{The mean radial magnetic field component for $r_0 = 0.9 {\rm R}$ ({\it red open circles and red curve}) and $r_0=0.7 {\rm R}$ ({\it blue dots and blue curve}) versus the daily cumulative sunspot area. The lines are the fitting by the fourth-order polynomial. }
\label{f1}
\end{figure}

\begin{figure} 
 \centerline{\includegraphics[width=0.5\textwidth,clip=]{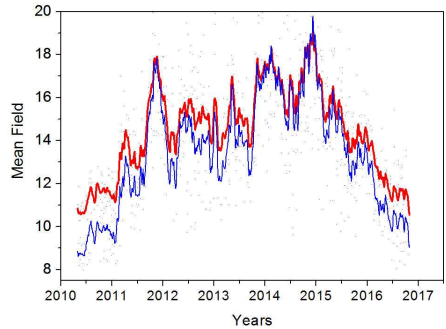}}
 \caption{Time dependence of the mean radial magnetic field component for $r_0 = 0.9 {\rm R}$ ({\it thick red line}) and $r_0=0.7 {\rm R}$ ({\it thin blue line}). The curves show  the 27-day running averaging.}
 \label{f2}
 \end{figure}

Since the direct measurements of the longitudinal-field component on the limb entail a number of errors, which introduce  a large uncertainty when making correction for the projection, we have performed the above calculations in a circle of a given radius ($r_0$) rather than over the whole hemisphere.
To start with, we compare the data obtained for $r_0 = 0.9 {\rm R}$ and $r_0 = 0.7{\rm  R}$, where ${\rm R}$ is the apparent solar radius, plotting the data versus the cumulative sunspot area (Figure~\ref{f1}) and time (Figure~\ref{f2}). One can see that the averaged data for both  $r_0$ are quite similar. At low solar activity, $r_0 = 0.9 {\rm R}$ gives a little larger mean values, which may be explained as the negative center-to-limb effect for weak background fields \citep{Ietal09}; however, this is unimportant for our statistical studies. So, below we are using $r_0 = 0.9 {\rm R}$ only.  

%

\section{Comparing Magnetic-Field Areas with Daily Cumulative Sunspot Areas}\label{comp}

We calculated $SB$, i.e. the daily areas on the Sun occupied by the magnetic fields larger than the threshold value $B_p$. The $SB$-values are calculated in the millionths of the visible hemisphere (m.v.h.) for $Bp$ from 0 to 1800 G. As a first step to finding the threshold magnetic field, we found the regression between $SB(B_p)$ and the cumulative sunspot area $Ss$

\begin{equation}
Ss= a(Bp)+b(Bp) SB (Bp),
\label{eq1}
\end{equation}
considering the regression coefficients $a$ and $b$ as functions of $B_p$. Since both $Ss$ and $SB$ are measured in the same units, we expect that the desired threshold value will correspond to $b \approx 1$ and $a \approx 0$. Indeed, this happens for $B_p = 550 - 1000$ G. In the same interval of $B_p$, the correlation between $Ss$ and $SB$ is the highest, which corroborates our expectations.

\begin{figure} 
 \centerline{\includegraphics[width=\textwidth,clip=]{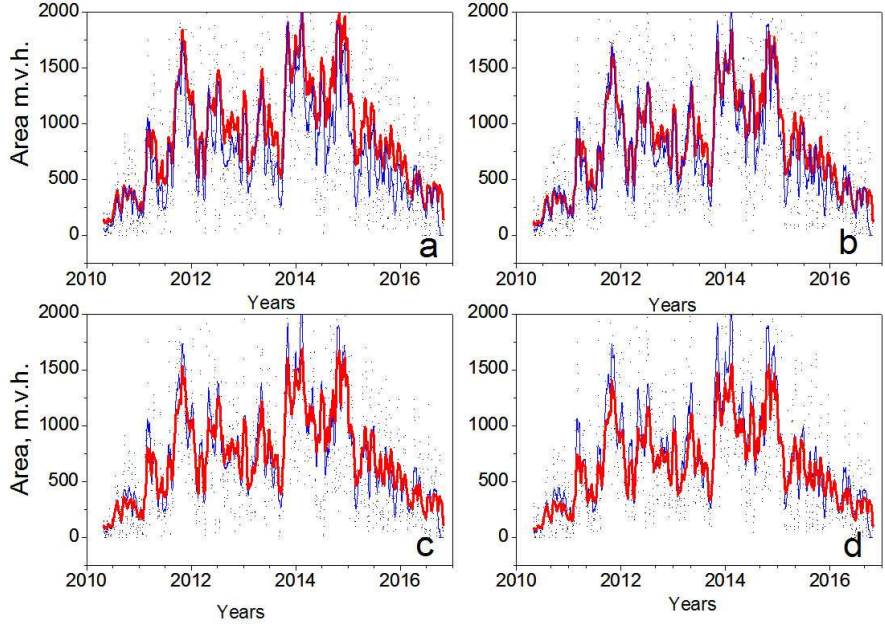}}
 \caption{$Ss$ ({\it thin blue line}) and the area bounded by the contour line for a given $B_p$ ({\it thick red line}) versus time. {\it Upper row: left} --- $B_p = 500$\,Mx\,cm$^{-2}$, {\it right} --- $B_p = 525$\,Mx\,cm$^{-2}$; {\it lower row: left} --- $B_p = 550$\,Mx\,cm$^{-2}$, {\it right} --- $B_p=575$\,Mx\,cm$^{-2}$. }
 \label{f3}
 \end{figure}

In order to establish the magnetic boundary of the penumbra more precisely, we show in Figure~\ref{f3} the cumulative sunspot area [$Ss$] (blue line) and the area bounded by a contour line (red line) for different threshold values of $B_p$ (500, 525, 550, and 575\,Mx\,cm$^{^{-2}}$) versus time. As seen from the figure, the red and blue lines are almost identical for $B_p = 525\,-\,575$\,Mx\,cm$^{-2}$.  

\begin{figure} 
 \centerline{\includegraphics[width=0.5\textwidth,clip=]{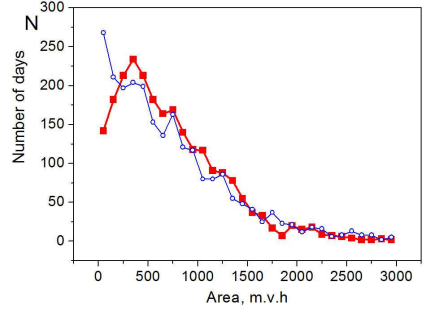}}
 \caption{Histograms (in the number [$N$] of days during the time intervals under discussion) for  the sunspots with the area bounded by $B_p=550$\,G ({\it thick red line and squares}) and the sunspots with the same cumulative areas ({\it thin blue line and open circles}).}
 \label{f4}
 \end{figure}

Additional verification is provided in Figure~\ref{f4}, where the histograms for the sunspots with the area bounded by $B_p=550$\,G and sunspots with the same cumulative areas are presented. The histograms are virtually identical except for the small sunspots. Interpretation of this difference is not clear. Perhaps, on the days of very low solar activity, when the total area was less than 150 m.v.h., the observers took into account the objects with a weak darkening (pores?), in which the fields were less than 550\,Mx\,cm$^{-2}$, and included their contribution to the cumulative area.  

Putting together all of the above, we arrive at the conclusion that, on average, the magnetic boundary of a sunspot as defined by the normal component of the magnetic field is 550\,G. The boundary is quite sharp, since at $B_p=500$\,G and $B_p=575$\,G, the agreement between the magnetic and visual data is obviously worse.

\section{Mean Values of the Magnetic Field in Sunspots} \label{aver}
 
\begin{figure} 
 \centerline{\includegraphics[width=0.5\textwidth,clip=]{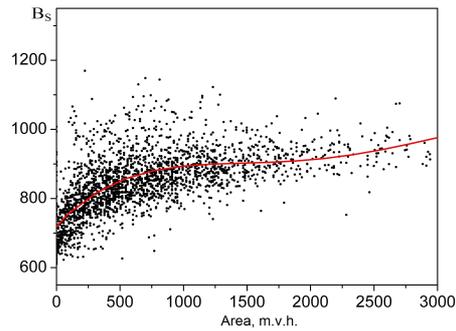}}
 \caption{Mean sunspot magnetic field versus the daily cumulative sunspot area. The curve is the fitting by the fourth-order polynomial.}
 \label{f5}
 \end{figure}

\begin{figure} 
 \includegraphics[width=0.45\textwidth,clip=]{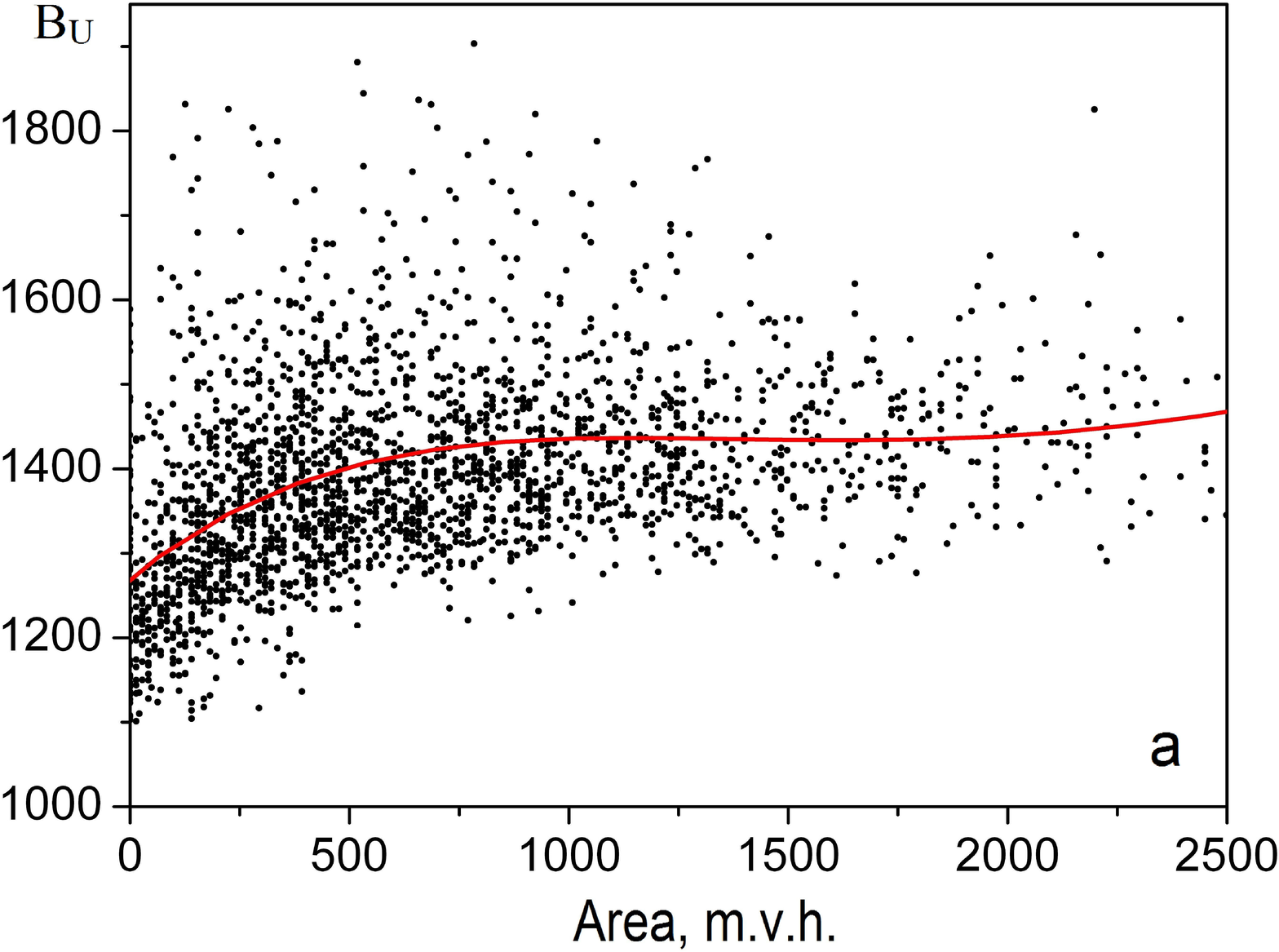}
 \includegraphics[width=0.45\textwidth,clip=]{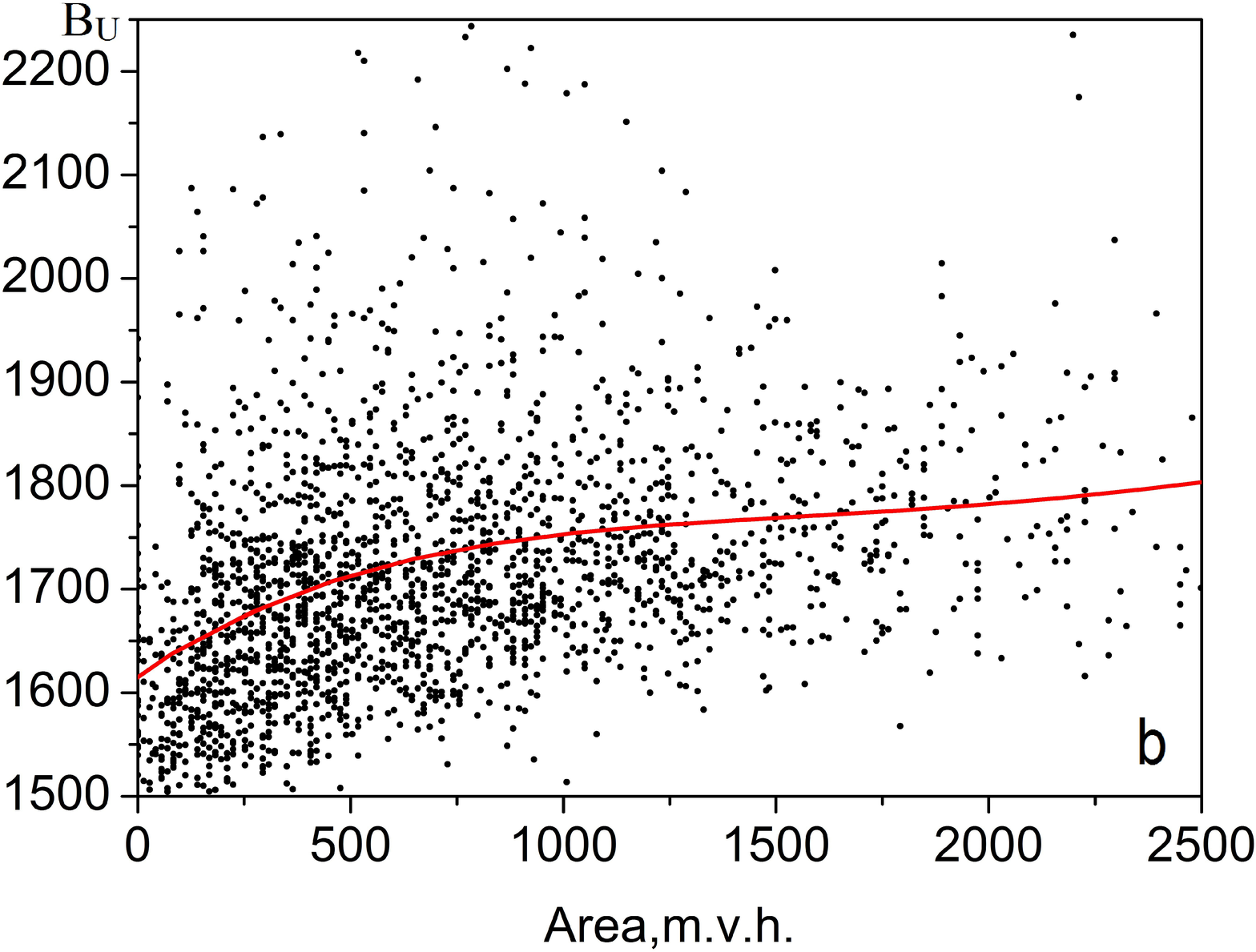}
 \caption{Mean sunspot magnetic field versus the daily cumulative sunspot area under the assumption that the umbra boundary corresponds to the magnetic field of 1100\,Mx\,cm$^{-2}$ ({\bf a}) and  1500\,Mx\,cm$^{-2}$ ({\bf b}). The curves are fitting by the fourth-order polynomial.}
 \label{f6}
 \end{figure}

Based on the threshold value of the sunspot magnetic  boundary, we can now estimate the mean magnetic field in sunspots (Figure~\ref{f5}). This proves to be about 750\,G for the smallest spots, becomes slightly larger (about 900\,G) on the days when the cumulative sunspot area reaches 500\,--\,2000 m.v.h, and increases somewhat more at higher solar activity. This value is smaller rather estimates discussed below; however, it should be borne in mind that the main contribution is made by the penumbra, which accounts for 80\,\%  of the sunspot area \citep{Betal14, BK19}. The photometric boundary of the umbra, i.e. the inner boundary of the penumbra is usually more diffuse than its outer boundary. Moreover, the relative area of the umbra  (15\,--\,30\%) depends on the sunspot size and varies in time (e.g., see \cite{Betal14} and references therein, as well as the discussion in Section \ref{intr} above). 
Based on the relative umbral area, one can admit that the boundary of the umbra  corresponds to the field value of 1100 G. The mean magnetic fields obtained under this assumption are shown in Figure~\ref{f6}a. Of course, if we accept the boundary according to \cite{Jetal15}, the field values in the umbra will be approximately 20\,--\,30\,\% larger. Figure~\ref{f6}b shows the mean magnetic field in the umbra under the assumption that the umbral boundary corresponds to the vertical magnetic field component as large as 2000\,Mx\,cm$^{-2}$. This is smaller than 2050\,G reported by \cite{N05, LW15, Netal16}. However, the latter estimate was obtained by averaging the maximum rather than the mean field values in the umbra.

\section{Estimating the Total Magnetic Flux}

\begin{figure} 
 \includegraphics[width=0.45\textwidth,clip=]{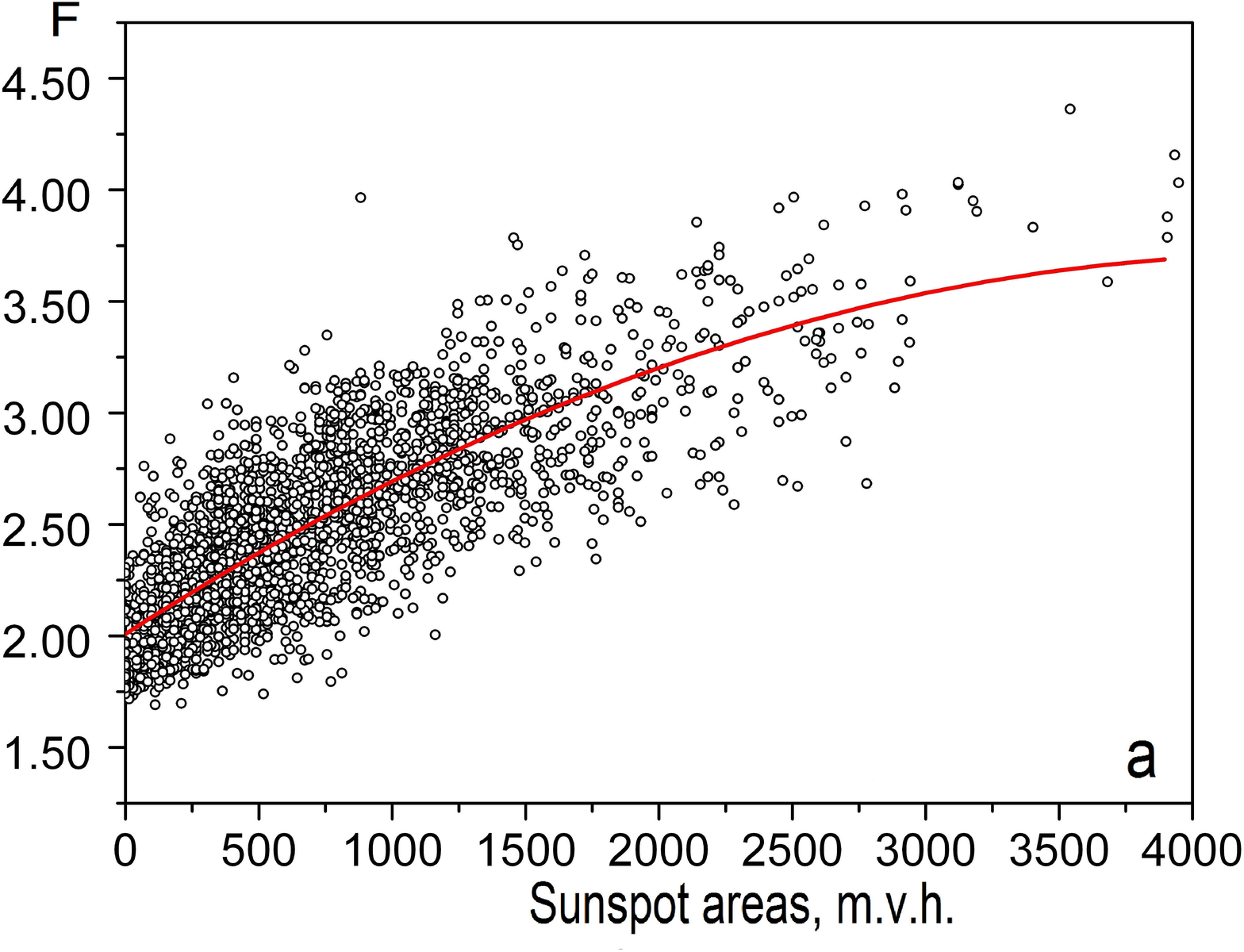}
 \includegraphics[width=0.45\textwidth,clip=]{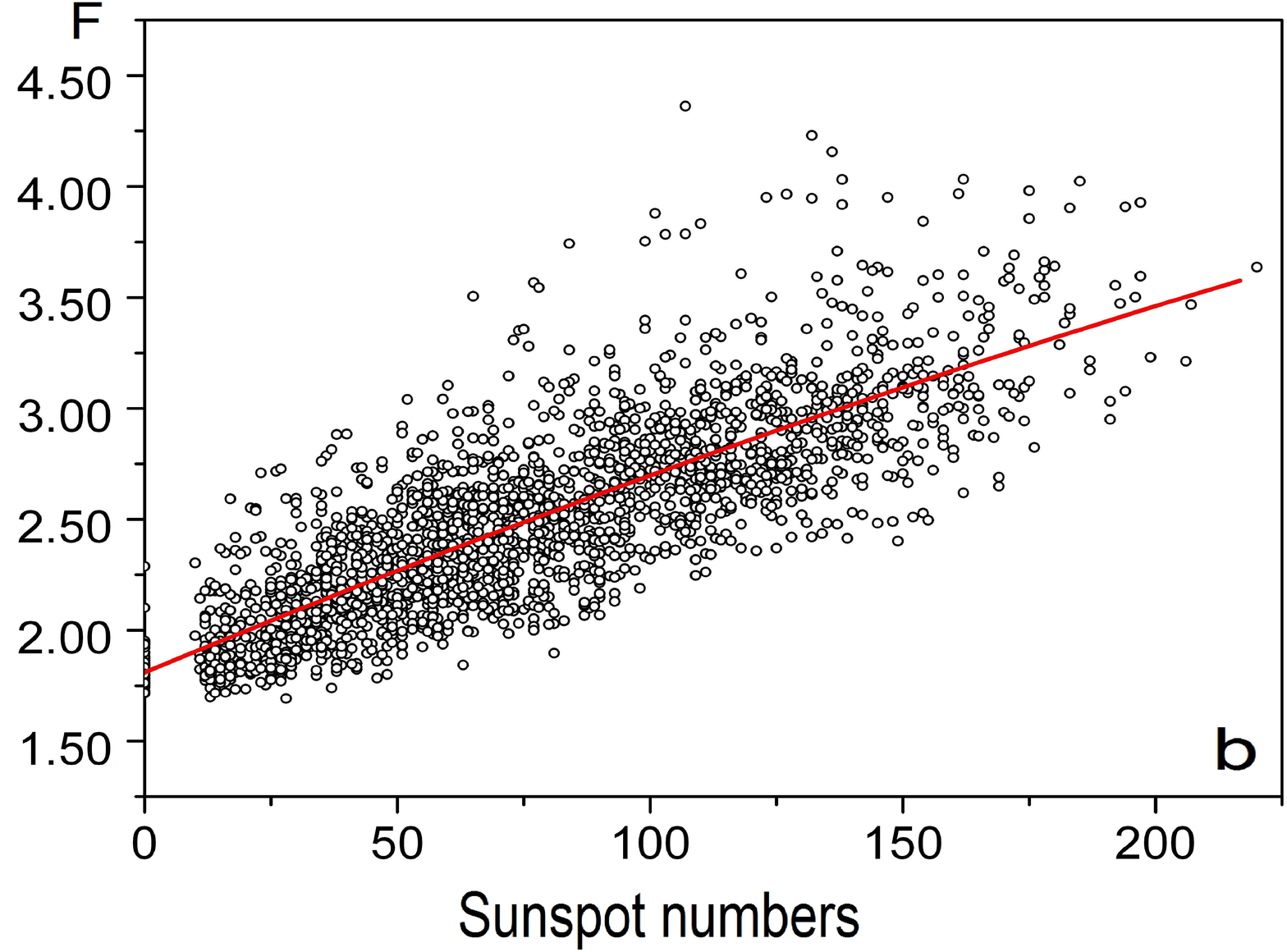}
 \caption{The total magnetic flux ({\bf a}) versus the cumulative sunspot area (the curves stand for best fitting $F= 2.00947 \times 10^{23}  + 7.71759 \times 10^{19} Ss - 8.74952 \times 10^{17} Ss^2$) and ({\bf b}) versus the sunspot number (the curves  stand for best fitting $F= 1.80896 \times 10^{23} + 9.50691 \times 10^{20} N - 6.23414 \times 10^{17} N^2$). Here $F$ is the magnetic flux in Mx\,$\times 10^{23}$, $N$ is the sunspot number, and $Ss$ is the cumulative sunspot area in m.v.h.}
 \label{f7}
 \end{figure}

\begin{figure} 
 \includegraphics[width=0.3\textwidth,clip=]{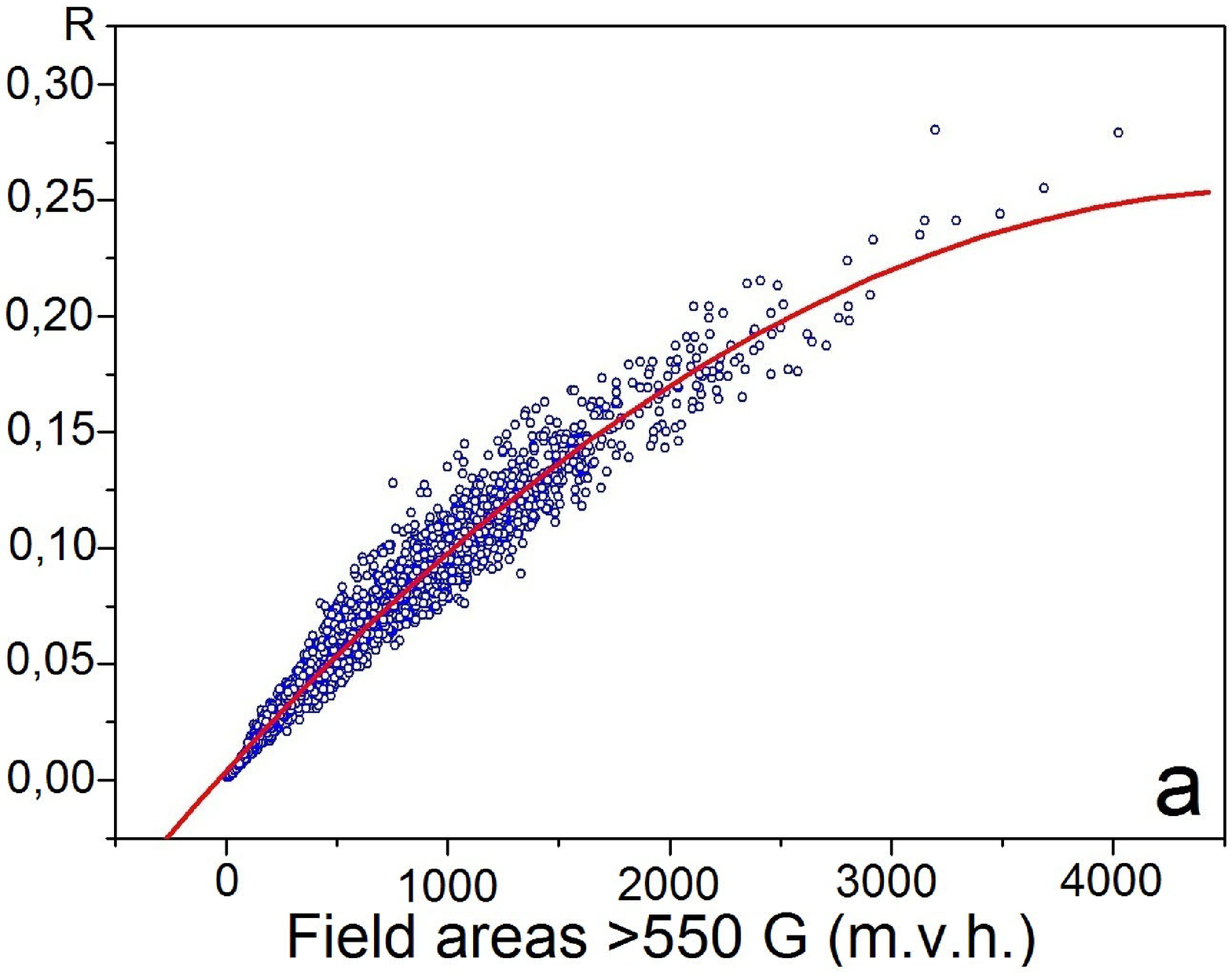}
 \includegraphics[width=0.3\textwidth,clip=]{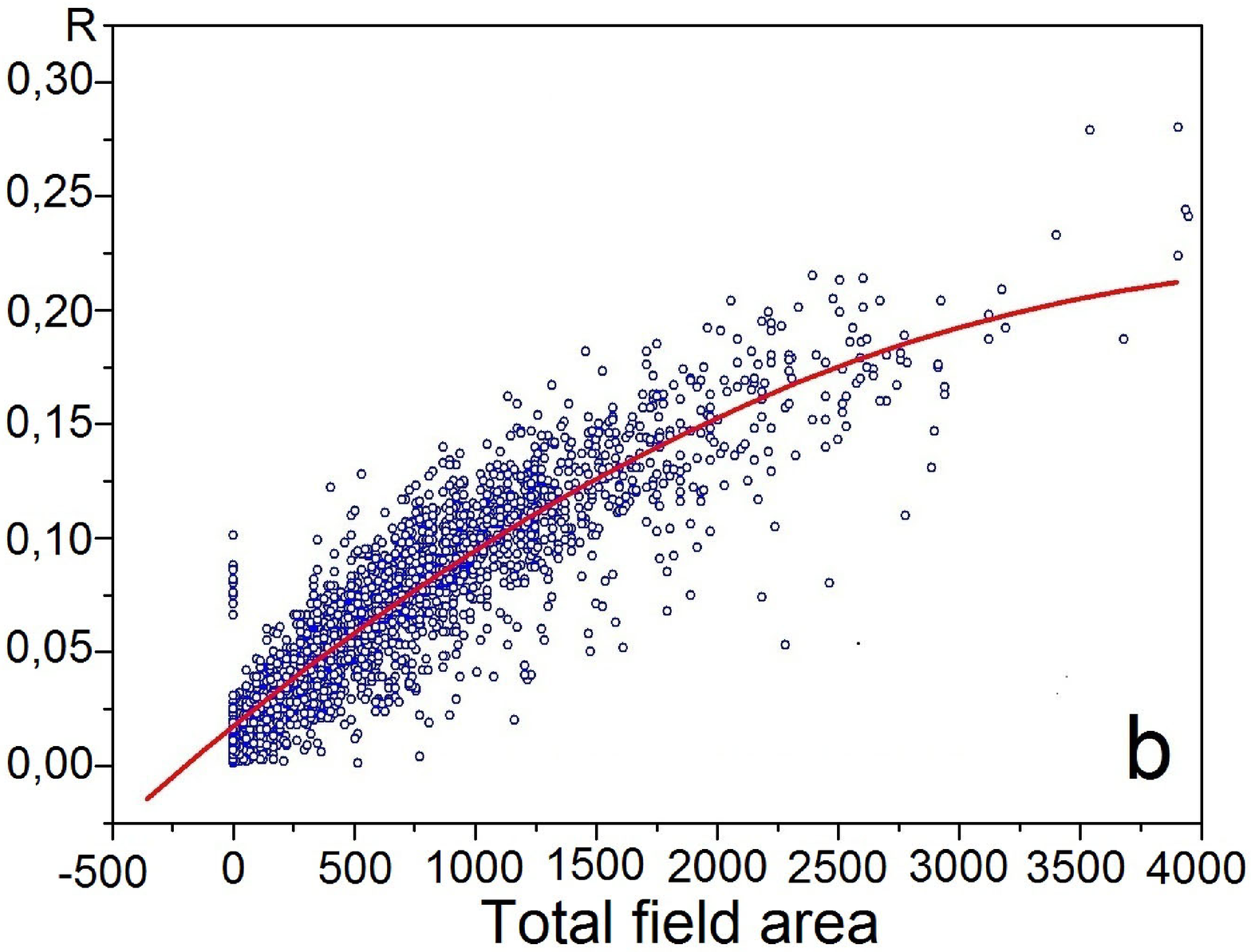}
 \includegraphics[width=0.3\textwidth,clip=]{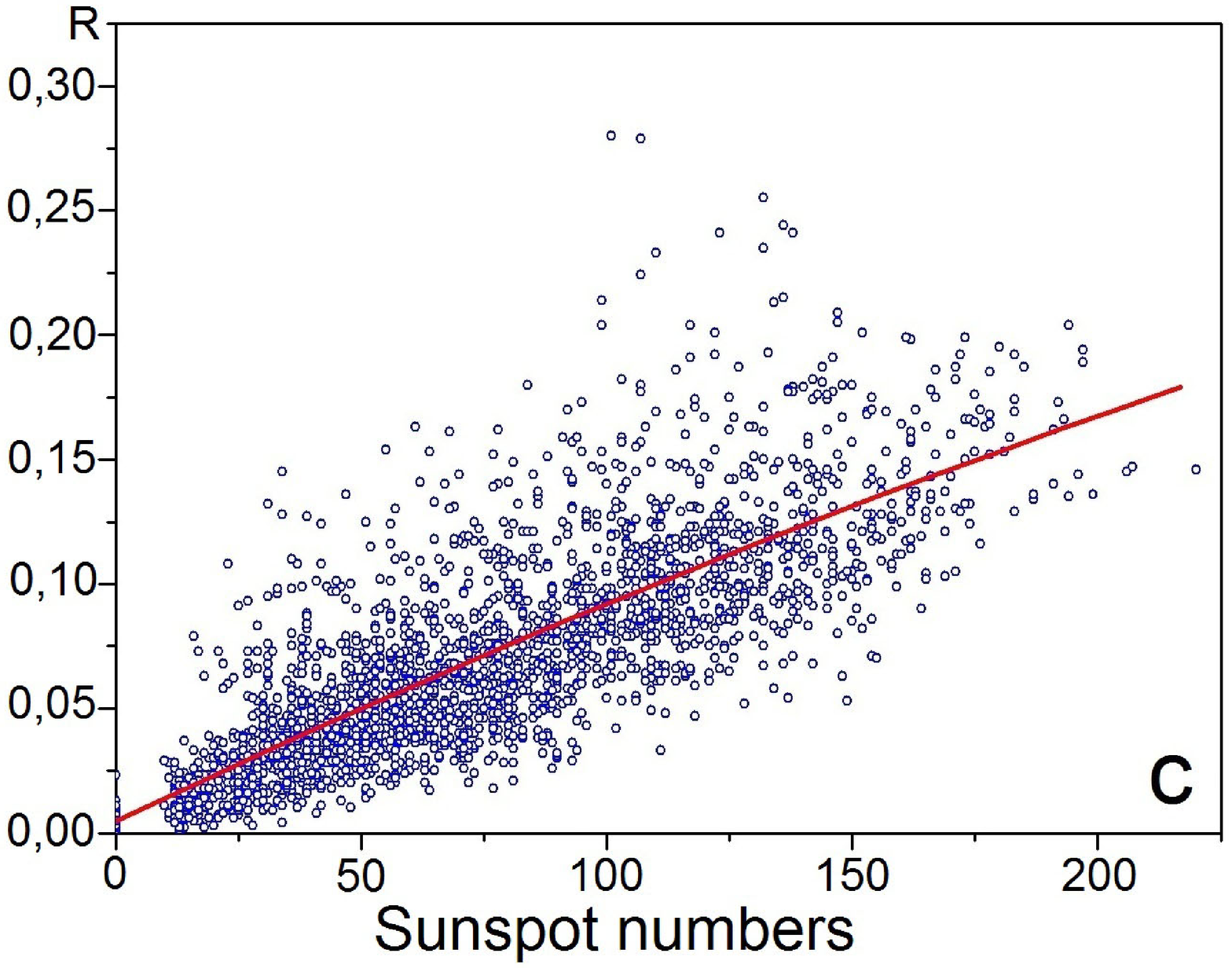}
 \caption{The ratio of the spot-magnetic flux to the total one versus {\bf a} - area occupied by magnetic field larger than 550\,G, {\bf b} - magnetic flux calculated for total sunspot area, {\bf c} - sunspot number.}
 \label{f8}
 \end{figure}

The results obtained above allow us to estimate the total magnetic flux and its dependence on local manifestations of the solar activity (Figure~\ref{f7}). For example, at $Ss = 1000$ m.v.h., the total magnetic flux is $2.693 \times 10^{23}$\,Mx, while the magnetic flux from sunspots of the same area is $2.49 \times 10^{22}$\,Mx only. With the transition to more active phases of the cycle, both the total flux and the flux from sunspots become larger, but the latter increases faster. The point is that the pure spotted flux is a relatively small part of the total flux, and the relationship between both fluxes is essentially non-linear (Figure~\ref{f8}). Therefore, the fitting formulas given in Figure~\ref{f7} ensure better estimates of the total magnetic flux than might be obtained taking into account the spot contribution alone.

\section{Conclusions and Discussion}

Based on the above results, we have arrived at the following conclusions:

In terms of the magnetic data, a sunspot can be described as a region, where the normal magnetic field component exceeds 525\,--\,550\,Mx\,cm$^{-2}$. The mean vertical magnetic field in a sunspot including the penumbra is about 900\,Mx\,cm$^{-2}$; the mean field in the umbra is about 1400\,--\,2000\,Mx\,cm$^{-2}$. These estimates depend little on the sunspot number as well as on the cumulative sunspot area and, therefore, they are almost independent of the phase of the solar cycle. It is to be borne in mind that we are talking here of the vertical magnetic field. The total magnetic field at the boundary of the penumbra can be three times larger. The mean magnetic field in the umbra depends substantially on the assumed field value at the inner boundary of the penumbra. 

In fact, it is not clear nowadays which of the magnetic-field components is responsible for the darkening at the spot boundary. This problem is additionally complicated by the fact that the magnetic-flux tube may be inclined relative to the normal to the solar surface. Therefore, it is not obvious whether in each spot, the contour line of the vertical-field component will accurately outline the photometric boundary. Here, we are dealing with subtle aspects of the relationship between the magnetic and temperature properties of a sunspot that are beyond the scope of our article. We intend to return to this issue in a separate article.

As for the total magnetic flux from the disc as a whole, it substantially and nonlinearly depends on the spottedness index.  This flux is about $2.0 \times 10^{23}$\,Mx even in the absence of sunspots and almost doubles at the cycle maximum. The contribution of sunspots is but a minor part of the total magnetic flux; however, it increases  by about a factor of ten  at the cycle maximum.

We should emphasize again that the particular characteristics of individual spots can differ greatly from those given in this work. Our task was to link together three independent databases, to identify the magnetic field values that agree best with the long-term photometric observations of sunspots, and allow  recalculation of data from one system to another.

It should be noted that the observed HMI values of B(LOS) in the sunspot umbra are most likely underestimated. In the umbra, where the 6173\,\AA \, line is in the strong-field regime, HMI only has five spectral wavelength points. Therefore, our estimates probably yield the lower value of the magnetic field in the umbra. It should be noted, however, that \cite{N05}, \cite{LW15}, and \cite{Netal16} give similar values. 

Of course, we understand that data obtained from B(LOS)-maps with Milne--Eddington inversions (HMI SHARP data) may not be consistent. There are disagreements
between them \citep{Dalda17}. This is an important question that can be solved by direct comparison of the data from these instruments for a selected number of particularly large sunspots.

Although our results seem to be specific to the Sun, they are absolutely necessary to fit the solar activity in the general context of stellar activity. Indeed, studying the solar activity in terms of the solar physics, we could limit ourselves to observing the distribution of temperatures over the visible solar disk in the same way as the first observers did. Dealing with stellar activity, however, we have to compare various tracers of the stellar and solar activity just because of observational constraints. For example, this problem inevitably arises when comparing solar flares with very powerful flares recently discovered on solar-type stars \citep[e.g.][]{Metal12}. If such flares occurred on the Sun, they would seriously threaten our technological civilization. Therefore, scientists engaged in the study of solar and stellar activity need to focus on the problem of superflares, which is directly related to the topics discussed above.

%

%
%

\begin{authorcontribution}
Statement of the problem mainly belongs to V.N.Obridko, numerical methods were mainly developed by I.M.Livshits, computations were mainly performed by B.D.Shelting, D.D.Sokoloff, M.M.Katsova contributed to discussion of results, M.M.Katsova was responsible for stellar aspects.
\end{authorcontribution}

 \begin{ethics}
\begin{conflict}
We have no conflict of interests.
 \end{conflict}

Data Availability
We have used data on the daily longitudinal magnetic field SDO/HMI for 2375 days from 1 May 2010 to 31 October 2016. The daily sunspot numbers were borrowed from WDC--SILSO, Royal Observatory of Belgium, Brussels sidc.oma.be/   silso/datafiles (version 2). The cumulative daily sunspot areas were taken from the NASA Web site solarscience.msfc.nasa.gov/greenwch.shtml.

Funding
V.N.Obridko, M.M.Katsova, D.D.Sokoloff acknowledge the support of Ministry of Science and Higher Education of the Russian Federation under the grant 075-15-2020-780,
V.N.Obridko thanks RFBR for the support with grant no. 20-02-00150,
D.D.Sokoloff acknowledges  the financial support of the Ministry of Education and Science of
the Russian Federation as part of the program of the Moscow Center for
Fundamental and Applied Mathematics under the agreement 075-15-2019-1621.

Acknowledgments 
We are grateful to the reviewer for a valuable remark.

 \end{ethics}
%
%


 
 \bibliographystyle{spr-mp-sola}
 \bibliography{sola_bibliography_example}  
\IfFileExists{\jobname.bbl}{} {\typeout{}
\typeout{****************************************************}
\typeout{****************************************************}
\typeout{** Please run "bibtex \jobname" to obtain} \typeout{**
the bibliography and then re-run LaTeX} \typeout{** twice to fix
the references !}
\typeout{****************************************************}
\typeout{****************************************************}
\typeout{}} 
%
%
%
%

\end{article} 
\end{document}